\documentclass[aps, a4paper, twocolumn, nofootinbib, showpacs, amsfonts, amsmath,amssymb]{revtex4}

\usepackage{graphicx}% Include figure files
\usepackage{dcolumn}% Align table columns on decimal point
\usepackage{bm}% bold math
\usepackage{hyperref}

\def\be{\begin{equation}}
\def\ee{\end{equation}}
\def\bea{\begin{eqnarray}}
\def\eea{\end{eqnarray}}

\def\pa{\partial}

\begin{document}

\title{Can UV meet IR in the Swiss cheese?}

\author{Madina Abilmazhinova}
\email{madina.abilmazhinova@nu.edu.kz }
\affiliation{Department of Physics, School of Sciences and Humanities, Nazarbayev University, Kabanbay Batyr 53, 010000 Astana, Kazakhstan}

\author{Diana Kulubayeva}
\email{diana.kulubayeva@nu.edu.kz }
\affiliation{Department of Physics, School of Sciences and Humanities, Nazarbayev University, Kabanbay Batyr 53, 010000 Astana, Kazakhstan}

\author{Hrishikesh Chakrabarty}
\email{hrishikesh.chakrabarty@nu.edu.kz}
\affiliation{Department of Physics, School of Sciences and Humanities, Nazarbayev University, Kabanbay Batyr 53, 010000 Astana, Kazakhstan}

\author{Daniele Malafarina}
\email{daniele.malafarina@nu.edu.kz}
\affiliation{Department of Physics, School of Sciences and Humanities, Nazarbayev University, Kabanbay Batyr 53, 010000 Astana, Kazakhstan}

\date{\today}

\begin{abstract}
 
We consider the embedding of regular black holes in an expanding universe and study how the ultraviolet modifications to the Schwarzschild geometry that regularize the black hole singularity affect the exterior universe's expansion rate. 
We consider several proposals for the regular black hole geometry and obtain the corresponding Friedmann equations for a universe filled only with dust and black holes. We show that different proposals have different implications which may be distinguished.
We then test the hypothesis that the UV corrections to the black hole geometry may be responsible for the current phase of accelerated expansion. To this aim we constrain the value of the regular black hole UV cutoff parameter from observations. Interestingly we find that the best fit is obtained by values of the parameter corresponding to regular horizonless compact objects.

\end{abstract}

\maketitle

\section{Introduction} \label{sec1}

Black holes do not exist in vacuum, they exist in an expanding universe filled with matter. We know that General Relativity (GR), our best theory that describes both black holes and the universe's expansion, has problems in the ultraviolet (UV) regime, as indicated by the generic occurrence of space-time singularities \cite{Hawking:1970zqf}, such as the ones at the end of collapse leading to the formation of black holes~\cite{Penrose:1969pc}. At the same time observations in the infrared (IR) regime, i.e. over cosmological distances, suggest the existence of a repulsive effect, named dark energy (DE), which permeates the universe at large scales and appears to be responsible for the universe's current acceleration. At present a viable explanation for DE is missing as is a viable modification of the field equations that allows for the resolution of the black hole singularity. There exist several theoretical proposals for addressing both problems but no definite answer (see for example \cite{Malafarina:2024qdz} for modern insights into gravitational collapse and \cite{Zhang:2025lam} for a review of the current status of dynamical dark energy models). Then it is worth wondering whether the two problems may be related. 
The relation between cosmological expansion and local physics has been studied for a long time, since the pioneering works of McVittie \cite{McVittie:1933zz}, and has recently found renewed interest (see \cite{Carrera:2008pi} for a recent review).
In the present article we explore how modifications of classical black holes in the UV may induce repulsive effects at large scales.

The idea of embedding a black hole geometry in an expanding universe has a long history. It was first considered by McVittie in 1933 who developed a metric that interpolates between the Schwarzschild solution at short distances and a Friedmann-Lemaître-Robertson-Walker (FLRW) solution at large distances \cite{McVittie:1933zz}. 
The McVittie metric has been considered in various contexts from detailed explorations of its properties \cite{Nolan:1998xs,Nolan:1999kk,Nolan:2014maa} to the relation with de Sitter, the $\Lambda$CDM and inhomogeneous models \cite{Kaloper:2010ec,Lake:2011ni, Landry:2012nv, Carrera:2009ve}.
More recently the McVittie solution has been considered in relation to the proposed cosmological mass scaling of supermassive black holes \cite{Farrah:2023opk, Lacy:2023kbb, Croker:2024jfg}. In \cite{Gaur:2023hmk}, it was shown how the effect does not occur for the original McVittie solution. On the other hand, in \cite{Cadoni:2023lum} it was shown that non-singular black hole masses may couple to cosmic expansion. These works underline the importance of the relation between the local physics of black holes for the large scale dynamics of the universe \cite{Cadoni:2023lqe}. A subset of the McVittie solution, known as Sultana-Dyer solution \cite{Sultana:2005}, has also been considered in the context of the embedding of black holes in an expanding universe (see for example \cite{Faraoni:2009uy}) while generalizations of the McVittie solution have also been studied \cite{Faraoni:2007es, Gao:2008jv}.

Another way to relate a black hole solution to an expanding universe is the so-called Swiss cheese model, which was originally developed by Einstein and Straus in 1945 \cite{Einstein:1945id}. The original Swiss cheese idea relies on the exact matching of a Schwarzschild black hole with an exterior FLRW dust universe. This model has been used to describe a black hole immersed in a dust universe \cite{schucking1954schwarzschildsche}, in a radiation filled universe \cite{hacyan1979radiation, reed1980comment} and in an inhomogeneous FLRW model \cite{lake1980local, bonnor2000generalizationof,Grenon:2009sx,Grenon:2011fs}.
More recently the Swiss cheese model has been considered as a potential new mechanism for cosmic acceleration driven by primordial black holes~\cite{Dialektopoulos:2025mfz}, or quantum corrections to black holes~\cite{Lewandowski:2022zce}.

Our aim in the present work is to use the Swiss cheese model to investigate whether proposed UV modifications to black hole geometries may have IR effects relevant to the universe's acceleration. The idea of relating the local physics of black holes to large scale expansion is intriguing and may have several potentially interesting implications \cite{Davidson:2012si, Faraoni:2024ghi}. However, it is also important to be able to relate such ideas to our current understanding of the universe's expansion as derived from observations. For this reason, we attempt to constrain the validity of the proposals with available cosmological data. This is an important step towards placing the interplay between black hole solutions and cosmological expansion on more solid footing.
Hence in the present paper we consider the matching conditions between different proposals for regular black hole geometries and an expanding universe in the Swiss cheese framework and 
analyze how they could be observationally distinguished by measuring the universe's expansion rate. 
While similar proposals can be found in the literature (see for example \cite{Torres:2026jht}) they generally lack observational constraints. In the present work we will attempt to restrict the allowed value of the parameter describing the UV cutoff of the regular black hole from cosmological observations in order to address the possibility that the UV corrections to black hole solutions that allow for the resolution of the central singularity may be responsible for the observed accelerated expansion rate of today's universe.

We will focus on well known regular black hole solutions such as the Hayward~\cite{Hayward:2005gi}\footnote{It must be noted that a solution formally identical to Hayward's had been previously obtained by Israel and Poisson in \cite{Poisson:1988wc}.}, Bardeen \cite{Bardeen}, Dymnikova \cite{Dymnikova:1992ux} and a recently proposed Asymptotically Safe solution \cite{Bonanno:2023rzk}.
For the observational constraints we will rely on available datasets such as the recent measurements of Baryon Acoustic Oscillation (BAO) by the Dark Energy Spectroscopic Instrument (DESI) \cite{DESI:2019jxc, DESI:2024mwx, DESI:2025zgx}, the Cosmic Chronometers (CC) measurements of the Hubble expansion rate \cite{Jimenez:2001gg, Borghi:2021rft} and the Type Ia Supernovae (SNe Ia) data from the Pantheon+ compilation \cite{Scolnic:2021amr}.

The paper is organized as follows: In section \ref{sec2} we discuss the most general matching conditions for two spherically symmetric space-times separated by a time-like spherical boundary. In section \ref{sec3} we restrict the attention to co-moving boundaries and analyze the induced dark energy component in the exterior universe as obtained from various proposals for regular black holes. Section \ref{sec4} deals with tests of the validity of the previously obtained cosmological models as a description of the current expansion rate of the universe. 
Finally in section \ref{sec5} we outline our results in the context of future observations.
Throughout the paper we make use of units such that $G=c=1$, unless otherwise stated.

\section{Swiss Cheese and junction conditions} \label{sec2}

We shall consider here the matching conditions between a static and spherically symmetric (regular) black hole solution and a FLRW universe. 
Since the Swiss cheese model is built upon the matching of two manifolds across a boundary we shall make use of the matching conditions in GR, which were originally investigated by Lanczos, Darmois and Lichnerowicz and later formalized in the currently most used form by Israel \cite{Israel:1966rt, Barrabes:1991ng}. The extension of the matching conditions to more general geometries such as radiating space-times was done in \cite{fayos1991matching,Fayos:1996gw}.
The matching conditions between Schwarzschild with linear stationary and axially symmetric perturbations and FLRW with arbitrary linear perturbations was also considered in~\cite{Mars:2008tq}.

It is useful to keep in mind that the same matching conditions apply when one considers a black hole for the interior and an expanding FLRW universe for the exterior, as we shall do here, as well as when one considers a collapsing interior and a black hole exterior (as it is done for example in the Oppenheimer-Snyder-Datt model~\cite{1939PhRv...56..455O, 1938ZPhy..108..314D}). 

Therefore, we shall take the black hole metric for the interior manifold $\mathcal{M}_-$, denoted by subscript $(-)$, in Schwarzschild coordinates $\{x_-^\mu\}=\{T,R,\theta,\phi\}$ as
\begin{equation}
    ds_{-}^2=-f(R)dT^2+\frac{dR^2}{f(R)}+R^2d\Omega^2,
\end{equation}
where $d\Omega^2$ is the line element on the unit 2-sphere and
\be 
f(R)=1-\frac{2M(R)}{R}.
\ee
Obviously Schwarzschild is obtained for $M=M_{\rm bh}$ being a constant.
For the matching surface $\Sigma_-$ we will take a dynamical spherical boundary radius $R_b(T)$ given implicitly by
\begin{equation}
    \Phi_{-}(x_-^\mu)=R-R_b(T)=0.
\end{equation}

For the exterior expanding universe, i.e. the exterior manifold $\mathcal{M}_+$ denoted by subscript $(+)$, we use co-moving coordinates $\{x_+^\mu\}=\{t,r,\theta,\phi\}$ for which the FRLW line element is
\begin{equation}
    ds_{+}^2=-dt^2+\frac{a^2}{1-kr^2}dr^2+r^2a^2d\Omega^2,
\end{equation}
and the boundary surface $\Sigma_+$ is given by a dynamical boundary radius $r_b(t)$ via
\begin{equation}
    \Phi_{+}(x_+^\mu)=r-r_b(t)=0.
\end{equation}

By restricting the metric on the boundary surfaces on both sides we obtain the three dimensional line elements
\bea 
ds^2_{\Sigma_-}&=& -\left[f(R_b)-\frac{1}{f(R_b)}\left(\frac{dR_b}{dT}\right)^2\right] dT^2 + R_b^2d\Omega^2, \\
ds^2_{\Sigma_+}&=& -\left[1-\frac{a^2}{1-kr_b^2}\left(\frac{dr_b}{dt}\right)^2\right] dt^2 + r_b^2a^2d\Omega^2.
\eea 
Since both line elements must describe the same boundary surface, i.e. $\Sigma_-=\Sigma_+=\Sigma$, the first matching conditions are obtained by imposing continuity of the metric functions across the surface. Notice that throughout we shall use Greek indices to denote four dimensional quantities, i.e. $\mu=0,1,2,3$ and Latin indices, i.e. $i,j$, to denote three dimensional quantities on $\Sigma$.

At this point, it is important to notice two things: First, in order to avoid an unnecessary proliferation of time coordinates, we will use the FLRW time as the time on the boundary and thus take the boundary surface as $\Sigma=\Sigma_+$ with coordinates $\{t,\theta,\phi\}$. However, one could also consider another co-moving time coordinate $\tau(t)$ in order to write the line element on the same boundary surface in the coordinates $\{y^i\}=\{\tau,\theta,\phi\}$ as 
\be 
ds^2_\Sigma=\gamma_{ij}dy^i dy^j=-d\tau^2+R_b(\tau)^2d\Omega^2.
\ee 
The two obviously coincide for a co-moving boundary with $r_b$ constant but need not coincide if $r_b$ is a function of $t$. Secondly, and related to the previous observation, if we do not impose the FLRW matching to be co-moving (i.e. $r_b={\rm const}.$) we will have one additional function to be determined from the matching conditions, namely $r_b(t)$. This may seem to introduce some additional freedom but at the same time we must keep in mind that the two conditions coming from matching of the second fundamental form will be independent in this case. Therefore we will have one additional function that must be obtained by solving an additional equation.

The matching conditions for the metric impose that $[g_{\theta\theta}]=0$ and $[g_{tt}]=0$, where we have used the notation $[A]=A_+-A_-$ to denote the jump of a quantity $A$ across $\Sigma$. Of course $[g_{\phi\phi}]=0$ is implied from $[g_{\theta\theta}]=0$ due to spherical symmetry. Then the first matching condition $[g_{\theta\theta}]=0$ allows us to express the boundary radius on the black hole side $R_b$ in terms of the boundary radius $r_b$ on the FLRW side and the scale factor as
\begin{equation}
R_b(T(t))=r_b(t)a(t).
\end{equation}
The second matching condition allows us to define the time coordinate on one side in terms of the time coordinate of the other side (up to an integration constant) and obtain $T(t)$ from
\begin{equation}
    \frac{dT}{dt}=\frac{1}{\sqrt{f(R_b)}}\left(1+\frac{(\dot{r}_ba+r_b\dot{a})^2}{f(R_b)}-\frac{a^2\dot{r}_b^2}{1-kr_b^2}\right)^{1/2},
\end{equation}
where we have defined $\dot{r}_b=dr_b/dt$, i.e. dotted quantities are used to denote derivatives w.r.t. the FLRW time $t$.

In order to obtain the matching for the second fundamental form we need the normal unit vectors to $\Sigma$ on both sides. These are defined as 
\be 
n_\mu^{\pm}=\frac{\pa_\mu \Phi_{\pm}}{\sqrt{-g^{\alpha\beta}\pa_\alpha \Phi_{\pm} \pa_\beta \Phi_{\pm}}}.
\ee 
Then from the $(-)$ side we get
\bea
    n_T^-&=&-\sqrt{\frac{1-kr_b^2}{\chi}}(\dot{r}_ba+r_b\dot{a}), \\
    n_R^-&=&\frac{1}{f}\sqrt{\frac{f\chi+(\dot{r}_ba+r_b\dot{a})^2(1-kr_b^2)}{\chi}},
\eea
and $n_\theta^-=n_\phi^-=0$. Where in the above we have defined
\be
\chi= 1-kr_b^2-a^2\dot{r}_b^2 .
\ee 

Similarly from the $(+)$ side we get
\bea
    n_t^+&=&-\frac{a\dot{r}_b}{\sqrt{\chi}}, \\
    n_r^+&=&\frac{a}{\sqrt{\chi}},
\eea
and $n_\theta^+=n_\phi^+=0$.
From the normal unit vectors we can evaluate the second fundamental form on $\Sigma$, which is defined as
\be 
K_{ij}=n_\sigma\left(\frac{\pa^2 x^\sigma}{\pa y^i \pa y^j }+\Gamma^\sigma_{\mu \nu}\frac{\pa x^\mu}{\pa y^i}\frac{\pa x^\nu}{\pa y^j}\right).
\ee 
Therefore in general we have two more matching conditions, namely $[K_{\theta \theta}]=0$ and $[K_{tt}]=0$. However, it is important to remember that in the case where the matching surface is geodesic (such as for example a co-moving constant boundary $r_b={\rm const}.$) the second condition should become equivalent to the first.

In our construction we obtain
\bea
    K_{\theta\theta}^{-}&=&\frac{r_ba\Psi}{\sqrt{\chi}}, \\
    K_{\theta\theta}^{+}&=&\frac{r_ba(1-kr_b^2)}{\sqrt{\chi}},
\eea
where we have defined $\Psi$ from
\be 
\Psi= 1-kr_b^2+ar_b\dot{a}\dot{r}_b.
\ee 
Using the relation $\Psi^2=f\chi+(1-kr_b^2)\dot{R}_b^2$, the matching condition $[K_{\theta \theta}]=0$ can then be reduced to the simple relation
\be \label{SFF-1}
f(R_b)=1-kr_b^2-r_b^2\dot{a}^2.
\ee 

For the last matching condition, after some tedious calculations, we obtain
\bea
    K_{tt}^{-}&=&\frac{\Psi\dot{\chi}-2\dot{\Psi}\chi}{2\dot{R}_b(1-kr_b^2)\sqrt{\chi}} ,\\
    K_{tt}^{+}&=&-\frac{a\ddot{r}_b+2\dot{a}\dot{r}_b}{\sqrt{\chi}}+\frac{a^2\dot{a}\dot{r}_b^3-kar_b\dot{r}_b^2}{(1-kr_b^2)\sqrt{\chi}},
\eea
which gives
\be \label{SFF-2}
[K_{tt}]=\frac{r_b\dot{r}_b\chi}{\dot{R}_b}\left(\frac{\ddot{a}}{a}-\frac{\dot{a}^2+k}{a^2}\right)=0.
\ee 

This last condition is valid for $R_b$ not constant and in the following we will consider a dynamical boundary radius from the black hole side, thus always setting $\dot{R}_b\neq 0$.

However, it is interesting to notice that one can also consider a stationary boundary on the black hole side, thus taking $R_b={\rm const}.$ In this case we have a static matching and the boundary on the FLRW side must be related to the scale factor via $r_b\dot{a}+a\dot{r}_b=0$.
Then it follows that $f=\chi$ and, if we consider a Schwarzschild exterior, for the matching of $K_{\theta\theta}$ we obtain
\be 
\frac{2M_{\rm bh}}{R_b^3}=\frac{\dot{a}^2+k}{a^2}=\frac{8\pi}{3}\rho={\rm const},
\ee
This first matching condition implies that we must have constant density. However this obviously is not the constant density Schwarzschild interior but it is a de Sitter universe with only a cosmological constant as the matter content. This matching should not be possible and in fact, as expected, it is easy to see that we can not satisfy the second matching condition, i.e. in the second Friedmann equation we do not obtain $p=-\rho$. We can't use Eq.~\eqref{SFF-2} since $\dot{R}_b = 0$ and instead we get
\be
[K_{tt}]=\frac{8\pi}{3}\rho-\frac{M_{\rm bh}}{R_b^3},
\ee
which can not be zero if the condition $[K_{\theta\theta}]=0$ holds, thus showing that the continuous matching is not allowed. However one could account for the pressure difference at the boundary arising from the second Friedmann equation by introducing a thin shell as it is done in gravastar models \cite{Mazur:2004fk}.  
Finally it is worth mentioning that if one considers an arbitrary dust space-time without a thin shell surface layer then the matching must always be performed across a co-moving boundary \cite{humphreys2012regular}, as we shall do in the following section.

\section{Co-moving boundary} \label{sec3}

As already noted by Lake and Grenon \cite{Grenon:2009sx, Grenon:2011fs}, if we wish to consider a co-moving boundary $r_b={\rm const.}$ but do not want to restrict ourselves to dust we must allow for the mass function $M$ not to be constant. Therefore in this section we shall focus the attention on $r_b={\rm const}.$ and consider what kind of cosmological models can be matched to a given regular black hole with $M(R)$. In this case, the condition \eqref{SFF-2} is automatically satisfied and from condition \eqref{SFF-1} we obtain the equation of motion as
\be \label{eom}
\frac{2M(R)}{r_b^3}=a(\dot{a}^2+k).
\ee 
In the case of Schwarzschild we have $M=M_{\rm bh}={\rm const}.$ and therefore we can define the contribution of the black hole mass to the density of a spherical region with boundary $r_b$ as
\be \label{rho_0} 
\rho_{\rm bh} = \frac{3M_{\rm bh}}{4\pi r_b^3} = {\rm const.}
\ee 
We can then obtain the usual first Friedmann equation for dust from the matching condition \eqref{eom} as
\be
\frac{8\pi}{3}\rho=\frac{8\pi}{3}\frac{\rho_{\rm bh}}{a^3}=\frac{\dot{a}^2+k}{a^2}.
\ee

In order to move to the corresponding setup for regular black holes, without loss of generality, we can always define the mass function as
\be 
M(R)=M_{\rm bh}\Big(1+h(R)\Big),
\ee 
where $M_{\rm bh}$ may be related to the mass parameter of the Schwarzschild limit while $h(R)$ describes the departures from the Schwarzschild geometry. The corresponding density obtained by $h$ on the FLRW side shall then be interpreted as a new fluid component induced by the strong field corrections to the black hole geometry. From Eq.~\eqref{SFF-1}, we obtain the first Friedmann equation as
\be\label{FRW}
\left(\frac{\dot{a}}{a}\right)^2 + \frac{k}{a^2} = \frac{8\pi}{3}\rho_* = \frac{8\pi}{3}\frac{\rho_{\rm bh}}{a^3}\Big(1+h(a)\Big),
\ee 
where $h(a)$ must be understood from $h(R_b)=h(r_ba)$ with $r_b$ constant.
Then the equation of state for the new fluid component $\rho_*$ can be defined as
\be \label{eos}
p_*=\omega_* \rho_*,
\ee 
with the equation of state parameter given by
\be\label{omega-h}
\omega_*(a) = -\frac{1}{3}\frac{h(a)_{,a}a}{1+h(a)}.
\ee

Notice that in order to obtain a de Sitter limit for the new fluid component we must have $h(a)$ such that $\omega_* \rightarrow -1$. For example, a de Sitter phase of acceleration in the early universe is typically obtained if $\omega_* \rightarrow -1$ for $a \rightarrow 0$.
In the following we shall consider some examples by choosing $h(R)$ and obtaining the corresponding FLRW and vice versa.

The simplest case we can consider for the interior metric is the (singular) Schwarzschild de Sitter geometry. This serves as a good test for the idea, since in this case the black hole mass $M_{\rm bh}$ and the cosmological constant $\Lambda$ decouple thus providing a universe with dust and dark energy where the dust component does not interact with the dark energy component. In this geometry the Schwarzschild part dominates at short distances and the de Sitter part dominates at large distances. The metric function for the Schwarzschild de Sitter space-time is
\be 
f(R)= 1 - \frac{2M_{\rm bh}}{R} - \frac{\Lambda}{3}R^2,
\ee 
for which we see that $M(R)=M_{\rm bh}+\Lambda R^3/6$.
Then on the FLRW we get
\be
    h(a) = \Lambda \frac{r_b^3a^3}{6M_{\rm bh}},
\ee
which leads to the following first Friedman equation
\be
    \left(\frac{\dot{a}}{a}\right)^2 + \frac{k}{a^2}=\frac{8\pi}{3}\left(\frac{\rho_{\rm bh}}{a^3} + \rho_\Lambda\right),
\ee
where
\be
    \rho_\Lambda = \frac{\Lambda}{8\pi} = -P_\Lambda.
\ee
Notice that if $M_{\rm bh}=0$ then there is no contribution to the dust-like matter content in the universe
%on the FRW side 
and the space-time is just pure de Sitter everywhere.

To avoid the occurrence of singularities at the center of black holes several proposals have been put forward that regularize the behavior at strong curvature by replacing the Schwarzschild core with a non singular de Sitter core. In the following we shall review some of most widely used solutions of this kind and apply them to the Swiss cheese framework to obtain the corresponding cosmological model.

\subsection{Regular black holes in non-linear electrodynamics}

A wide class of regular black holes can be obtained in GR coupled to a theory of non-linear electrodynamics as proposed by Fan \& Wang in \cite{Fan:2016hvf}. The mass function for this family of solutions is
\be \label{nled}
M(R)=M_{\rm bh}\frac{R^\mu}{(R^\nu+q_{ned}^\nu)^{\mu/\nu}} ,
\ee 
where $q_{ned}$ is related to the non-linear electrodynamics charge. Notice that we must require $\mu\geq 3$ in order for the black hole solution to be regular and we can retrieve known solutions for particular values of $\mu$ and $\nu$. For example, for $\mu=-1$, $\nu=1$ one obtains Maxwell's electrodynamics and the (singular) Reissner-Nordstrom black hole, for $\mu=\nu=3$ one retrieves the Hayward black hole~\cite{Poisson:1988wc,Hayward:2005gi} while $\mu=3$ and $\nu=2$ gives the Bardeen black hole~\cite{Bardeen}.

Following the procedure outlined above we can then match the black hole solution to a cosmological model with
\be
    h(a) = -1+\frac{ a^\mu}{\left( a^\nu + Q_{ned}^\nu \right)^{\mu/\nu}},
\ee
where we have defined $Q_{ned}=q_{ned}/r_b$ as the dimensionless charge parameter. The first Friedmann equation is given by Eq.~\eqref{FRW} with
\begin{align}
    \rho_{*}(a) = \frac{\rho_{\rm bh} a^{\mu-3}}{\left( a^\nu + Q_{ned}^\nu \right)^{\mu/\nu}},
\end{align}
and the equation of state parameter becomes 
\begin{align} 
\omega_{\rm *}(a)=-\frac{\mu}{3}\frac{ Q_{ned}^\nu}{a^\nu + Q_{ned}^\nu}.
\end{align} 
For large $a$ the equation of state becomes dust-like while remaining negative at all times. The de Sitter equation of state is recovered for small $a$ when $\mu=3$ (thus both the Hayward as well as the Bardeen black hole have a de Sitter limit). However, as we shall see later, in a universe filled with black holes and matter, it is possible to find values of $Q_{ned}$ and $M_{\rm bh}$, for which the component of the fluid given by this kind of regular black holes is still causing acceleration today.

\subsection{Dymnikova regular black hole}
Another widely studied regular black hole was obtained by Dymnikova  \cite{Dymnikova:1992ux}. This solution is obtained by relating a vacuum energy de Sitter core at small distances to a Schwarzschild vacuum solution at large distances. The mass function is
\be 
M(R)=M_{\rm bh}\Big(1-e^{-R^3/(2M_{\rm bh} q_D)}\Big),
\ee 
which gives
\begin{align} 
    h(a) = -e^{-r_b^3 a^3/(2M_{\rm bh} q_D)},
\end{align}
where $q_D$ is a parameter related to the characteristic length scale of the de Sitter core. Then the matching condition Eq.~\eqref{SFF-1} gives the Friedmann equation \eqref{FRW} with
\begin{align}
    \rho_*(a) = \frac{\rho_{\rm bh}}{a^3} \left( 1 - e^{- a^3/ \tilde{q}_D} \right),
\end{align}
where $\tilde{q}_D = 8\pi\rho_{\rm bh} q_D/3$.

The pressure is again given by Eq.~\eqref{eos} with
\begin{align}
    \omega_*(a) = \frac{a^3}{\tilde{q}_D(1-e^{a^3/\tilde{q}_D})}.
\end{align}
Similar to the previous case, we have $\omega_*<0$ and $\omega_*\rightarrow 0$ for $a$ large while the de Sitter equation of state is recovered for $a$ small.

\subsection{Regular black hole in Asymptotic Safety}
We shall now consider a newly proposed regular black hole in Asymptotic Safety \cite{Bonanno:2023rzk} for the interior solution. This solution was obtained from gravitational collapse\footnote{See for example \cite{Bonanno:2020fgp} for a compact object obtained within the same framework.} within the Markov-Mukhanov effective action \cite{Markov:1985py} by using the Bonanno-Reuter renormalization group approach \cite{Reuter:1996cp} to obtain a variable gravitational coupling in Asymptotic Safety \cite{Bonanno:2000ep}.
It is worth mentioning that constraints on Swiss cheese cosmologies in Asymptotic Safety have been investigated in a different context in  \cite{Anagnostopoulos:2018jdq,Anagnostopoulos:2022pxa}. For the line-element obtained in \cite{Bonanno:2023rzk} we have the following mass function
\be 
M(R)=\frac{R^3}{6q_{AS}}\log\left(1+\frac{6M_{\rm bh}q_{AS}}{R^3}\right),
\ee 
from which we get
\begin{align}
    h(a) = - 1 + \frac{r_b^3 a^3}{6M_{\rm bh} q_{AS}}\log\left( 1 + \frac{6M_{\rm bh} q_{AS}}{r_b^3a^3}\right), 
\end{align}
where $q_{AS}$ is a free dimensional parameter that sets the scale for UV cutoff of Asymptotic Safety. The matching condition in Eq.~\eqref{SFF-1} gives Eq.~\eqref{FRW} with
\begin{align}
    \rho_*(a) = \frac{\rho_{\rm bh}}{ \tilde{q}_{AS}}\log\left( 1 + \frac{ \tilde{q}_{AS} }{a^3} \right),
\end{align}
where $\tilde{q}_{AS} = 8\pi \rho_{\rm bh} q_{AS}$ is a constant.
The equation of state \eqref{eos} then has 
\be
    \omega_*(a) = -1 + \frac{\tilde{q}_{AS}}{\left(a^3+\tilde{q}_{AS}\right)\log\left( 1 + \tilde{q}_{AS}/a^3 \right)}.
\ee
Again, like in the previous cases, $\omega_*$ is negative, goes to $0$ for large $a$ and goes to $-1$ for small $a$.

\subsection{Dynamical dark energy inspired black hole}

Finally, we may also follow the opposite procedure by choosing a cosmological model and inferring the corresponding black hole mass function. In this case it makes sense to choose a dynamical dark energy content in alignment with current experimental constraints. 
The simplest example is given by a dynamical dark energy component governed by the generic (Chevallier-Polarski-Linder) CPL parametrization \cite{Chevallier:2000qy,Linder:2002et}, also known as $\omega_0 \omega_a$ parametrization. This is a phenomenological model that needs to be understood as a linear approximation for the dynamical dark energy close to the present time. Then the equation of state parameter is
\be
    \omega_*(a) = \omega_0 + \omega_a(1-a),
\ee
where $\omega_0$ and $\omega_a$ are two constants that need to be constrained from observations. Using Eq.~\eqref{omega-h}, we can integrate to find $h(a)$ as
\be
    1+h(a) = \frac{c e^{3\omega_a a}}{a^{3(\omega_0+\omega_a)}},
\ee
where $c$ is a dimensionless integration constant. It is important to notice that, since the CPL model is a phenomenological parametrization of the DE equation of state close to today, it is not useful to build a mass function $M(R)$ for a black hole, as the resulting geometry may not be valid for small $R$ or even close to the horizon.
However, we may retrieve the values of $\omega_0$ and $\omega_a$ in the CPL parametrization expressed in terms of the regular black hole parameter for the models considered above. In fact from Eq.~\eqref{omega-h} we get
\bea 
\omega_0&=&\omega_*(1)=-\frac{1}{3}\frac{h_{,a}(1)}{1+h(1)} ,\\
\omega_a&=&\left.\frac{d\omega_*}{da}\right|_{a=1}=\omega_0\left(1+3\omega_0+\frac{h_{,aa}(1)}{h_{,a}(1)}\right).
\eea 

\section{Phenomenological implications for the late universe}\label{sec4}

\begin{table*}[t]
\centering

\begingroup

\setlength{\tabcolsep}{4pt} % Default value: 6pt
\renewcommand{\arraystretch}{1.7} % Default value: 1

\begin{tabular}{c|w{c}{2cm}|w{c}{2.2cm}|c|c}
\hline
Model & $H_0 (\rm km/s/Mpc)$ & $\Omega_{\rm m}$ & Charge & $r_d (\rm Mpc)$  \\ \hline
$\Lambda$CDM & $68.22 \pm 0.48$ & $0.309 \pm 0.008$ & $-$ & $147.08 \pm 0.24$ \\
$w_0w_a$CDM & $66.87\pm0.55$ & $0.303^{+0.016}_{-0.026}$ & $-w_0=0.77\pm0.04$, $w_a=-0.42^{+0.42}_{-0.35}$ & $147.09\pm0.24$  \\
Hayward & $66.89\pm0.56$ & $0.301\pm0.008$ & $Q_{ned}>1.320$ & $147.09\pm0.24$ \\
Bardeen & $66.92\pm0.56$ & $0.297\pm0.008$ & $Q_{ned}>1.598$ & $147.09\pm0.24$ \\
Dymnikova & $66.84\pm0.56$ & $0.302\pm0.008$ & $Q_D = 1.112^{+0.067}_{-0.063}$ & $147.09\pm0.24$  \\
AS & $66.96\pm0.56$ & $0.287\pm0.009$ & $Q_{AS}>5.073$ & $147.09\pm0.24$ \\ \hline
\end{tabular}
\endgroup

\caption{
Median values and associated uncertainties for the inferred parameters from the MCMC analysis for $\Lambda$CDM, $w_0w_a$CDM and CCRBH DE models.  
For Hayward, Bardeen and AS CCRBH, the charge is reported as $95\%$ lower bound, because the upper limit is unconstrained from late time dataset. }
\label{tab1}
\end{table*}

As we have seen in the previous section, cosmologically coupled regular black holes (CCRBH) can act as an additional matter content of the universe contributing to the expansion rate. This additional content also posits equations of state that can mimic that of the dark energy component. Therefore, it would be natural to test this kind of proposal against observations. In this section, we consider four CCRBH dark energy models and confront them with late time cosmological data to put bounds on the values of the UV cutoff parameter. For simplicity, we assume the content of the universe to be made only of matter (comprising both baryonic and dark matter) and black holes, with the matter density $\rho_{\rm m}$ described by a dust-like equation of state and the black holes contributing to an additional energy density $\rho_*$ as described in the previous section.

\subsection{Model setup}

Let us first set up the models. In the following, we shall assume a flat universe, i.e. $k = 0$ and neglect any contribution to the expansion rate due to radiation. Hence we will consider a universe filled with matter ($\rm m$) and black holes ($\rm bh$).

\begin{itemize}
    \item[(a)] {\bf Hayward:}
    The Hayward regular black hole \cite{Hayward:2005gi} can be obtained from the wider class of regular black holes in nonlinear electrodynamics in Eq. \eqref{nled} by setting $\mu=\nu=3$ \cite{Fan:2016hvf}. The resulting Friedmann equation that governs the late time dynamics of the universe can be written as
    \begin{align}
        \frac{H^2}{H_0^2} = \frac{\Omega_{\rm m}}{a^3} + \frac{\Omega_{\rm bh}}{a^3+Q_{ned}^3}.
    \end{align}
    Here $H=\dot{a}/a$ is the Hubble parameter, $H_0$ is the value of the Hubble parameter today, $\Omega_{\rm m}$ is the density parameter of pressureless dust, $\Omega_{\rm bh} = \rho_{\rm bh}/\rho_{\rm c}$ is the density parameter for the black hole content with $\rho_{\rm c} = 3H_0^2/8\pi$ being the density of the universe today. 
    In the following, we enforce normalization condition $H = H_0$ at the present day, i.e. at $a = 1$, so that $H / H_0 = 1$ today and we get
    \begin{align}
        \Omega_{\rm bh} = (1-\Omega_{\rm m})(1+Q_{ned}^3).
    \end{align}
    Finally we express the Friedmann equation in terms of the redshift $z=1/a-1$ as
    \begin{align}\label{eq-fe-hay}
        \frac{H^2}{H_0^2} = \Omega_{\rm m}(1+z)^3 + \frac{\Omega_{\rm bh}(1+z)^3}{1+Q_{ned}^3(1+z)^3}.
    \end{align}

    \item[(b)] {\bf Bardeen:} Similarly the Bardeen black hole \cite{Bardeen} is obtained from Eq. \eqref{nled} for $\mu = 3$ and $\nu = 2$. The resulting Friedmann equation driving the late time expansion is given by
    \begin{align}
        \frac{H^2}{H_0^2} = \frac{\Omega_{\rm m}}{a^3} + \frac{\Omega_{\rm bh}}{\left(a^2+Q_{ned}^2\right)^{3/2}},
    \end{align}
    where $Q_{ned}$ is the dimensionless charge parameter of the cosmologically coupled Bardeen black hole. After enforcing the normalization $\Omega_{\rm bh}$ can be written as 
    \begin{align}
        \Omega_{\rm bh} = (1-\Omega_{\rm m})(1+Q_{ned}^2)^{3/2},
    \end{align}
    and the Friedmann equation in terms of the redshift becomes
    \begin{align}\label{eq-fe-bar}
        \frac{H^2}{H_0^2} = \Omega_{\rm m}(1+z)^3 + \frac{\Omega_{\rm bh}(1+z)^3}{(1+Q_{ned}^2(1+z)^2)^{3/2}}.
    \end{align}

    \item[(c)] {\bf Dymnikova:} The first Friedmann equation for an universe with Dymnikova \cite{Dymnikova:1992ux} CCRBH can be written as
    \begin{align}
        \frac{H^2}{H_0^2} = \frac{\Omega_{\rm m}}{a^3} + \frac{\Omega_{\rm bh}}{a^3}\left( 1 - e^{-a^3/\Omega_{\rm bh} Q_D} \right),
    \end{align}
    where $Q_D = 8\pi q_D \rho_{\rm c}/3$. From the normalization condition, similarly to the two earlier models, we find $\Omega_{\rm bh}$ as
    \begin{align}
        \Omega_{\rm bh} = \frac{1-\Omega_{\rm m}}{1 +  W\left( \varphi \right)(1-\Omega_{\rm m}) Q_D },
    \end{align}
    where $W(\varphi)$ is a Lambert W function and
    \begin{align}
        \varphi = - \frac{e^{-1/(1-\Omega_{\rm m})Q_D}}{(1-\Omega_{\rm m})Q_D}.
    \end{align}
    Finally we express the Friedmann equation in terms of the redshift
    \be 
    \frac{H^2}{H_0^2} =(1+z)^3\left[\Omega_{\rm m} + \Omega_{\rm bh}
        \left( 1 - e^{-1/\Omega_{\rm bh} Q_D(1+z)^3} \right)\right].
    \ee
    
    \item[(d)] {\bf Asymptotic Safety:} The first Friedmann equation for Asymptotically Safe \cite{Bonanno:2023rzk} CCRBH along with a pressureless matter  can be written as
    \begin{align}
        \frac{H^2}{H_0^2} = \frac{\Omega_{\rm m}}{a^3} + \frac{1}{Q_{\rm AS}}\log\left( 1 + \frac{\Omega_{\rm bh}Q_{\rm AS}}{a^3} \right),
    \end{align}
    where $Q_{\rm AS} = 8\pi q_{AS} \rho_{\rm c}$ is the dimensionless charge parameter.
    Enforcing the normalization, we find $\Omega_{\rm bh}$ as 
    \begin{align}
        \Omega_{\rm bh} = \frac{1}{Q_{\rm AS}}\left(e^{(1-\Omega_{\rm m})Q_{\rm AS}}-1 \right),
    \end{align}
    and in terms of the redshift, the first Friedmann equation becomes
    \begin{align}\label{eq-fe-as}
        \frac{H^2}{H_0^2} = &\Omega_{\rm m}(1+z)^3 +\frac{\log\left[ 1 + \Omega_{\rm bh}Q_{\rm AS}(1+z)^3 \right]}{Q_{\rm AS}}.
    \end{align}
\end{itemize}

In the following we will test and put constraints on the parameters of these four cosmological models with late time cosmological data. Notice that when the extra charge parameters of all these models, $Q_{ned}, Q_D$ and $Q_{AS}$ vanishes, the universe becomes only matter dominated, i.e. no dark energy, and the black holes contribute to the dust-like content. On the other hand at low redshifts the Hayward, Bardeen and Asymptotically Safe CCRBH dark energy models become indistinguishable from the $\Lambda$CDM. This can be easily seen from the corresponding Friedmann equations \eqref{eq-fe-hay}, \eqref{eq-fe-bar} and \eqref{eq-fe-as}. At low redshifts, for higher values of the charge parameter, the CCRBH dark energy term in the Friedmann equation becomes constant. Therefore, since we are dealing with low redshift data, we can put only lower bounds on the extra charge parameter of these CCRBH dark energy models. On the other hand, the Dymnikova CCRBH dark energy model doesn't have such a feature and it is possible to constrain the parameter at both ends.      

We investigate here the possibility to constrain the free parameters of the four models above, i.e. $\{ H_0, \Omega_m, Q_{ned} \}$ for Hayward, $\{ H_0, \Omega_m, Q_{ned} \}$ for Bardeen, $\{ H_0, \Omega_m, Q_D \}$ for Dymnikova and $\{ H_0, \Omega_m, Q_{AS} \}$ for Asymptotic Safety. In particular, we would like to put bounds on the UV cutoff charge for each CCRBH. For this purpose, we perform a parameter inference procedure on our models by using a publicly available sampler \texttt{emcee} \cite{emcee} to implement Markov Chain Monte Carlo (MCMC) procedure. A version of our code can be found at the public repository \texttt{jla$\_$ccrbh}\footnote{\href{https://github.com/rishid8/jla_ccrbh}{https://github.com/rishid8/jla$\_$ccrbh}}. We check the convergence of our MCMC chains using Gelman \& Rubin ${\rm R}-1$ parameter \cite{Gelman:1992zz} and we consider the convergence condition to be met for our chains when ${\rm R} - 1 < 0.01$.

\begin{figure*}[]
   \begin{center}
        \includegraphics[width=8cm]{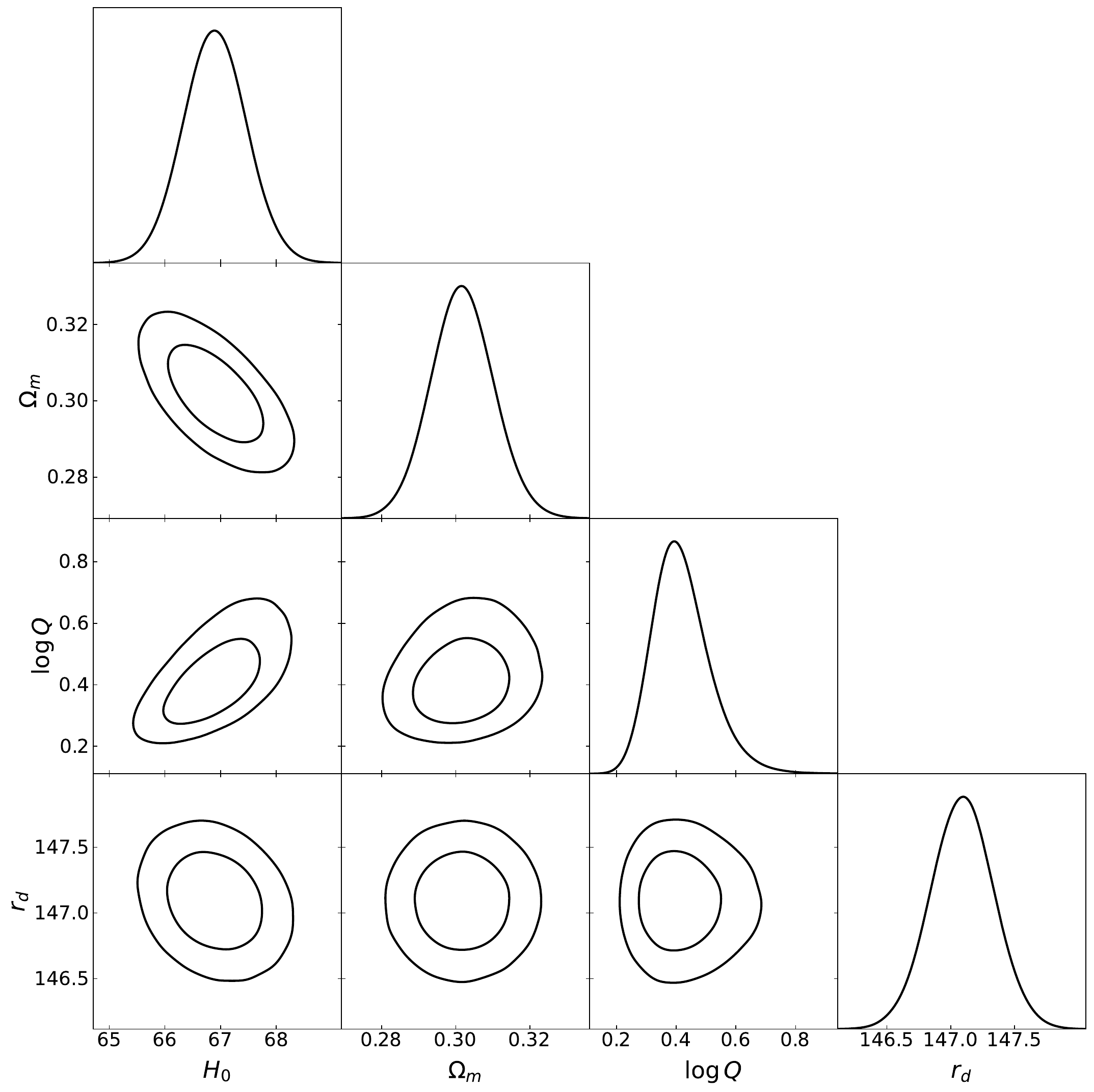}
        \includegraphics[width=8cm]{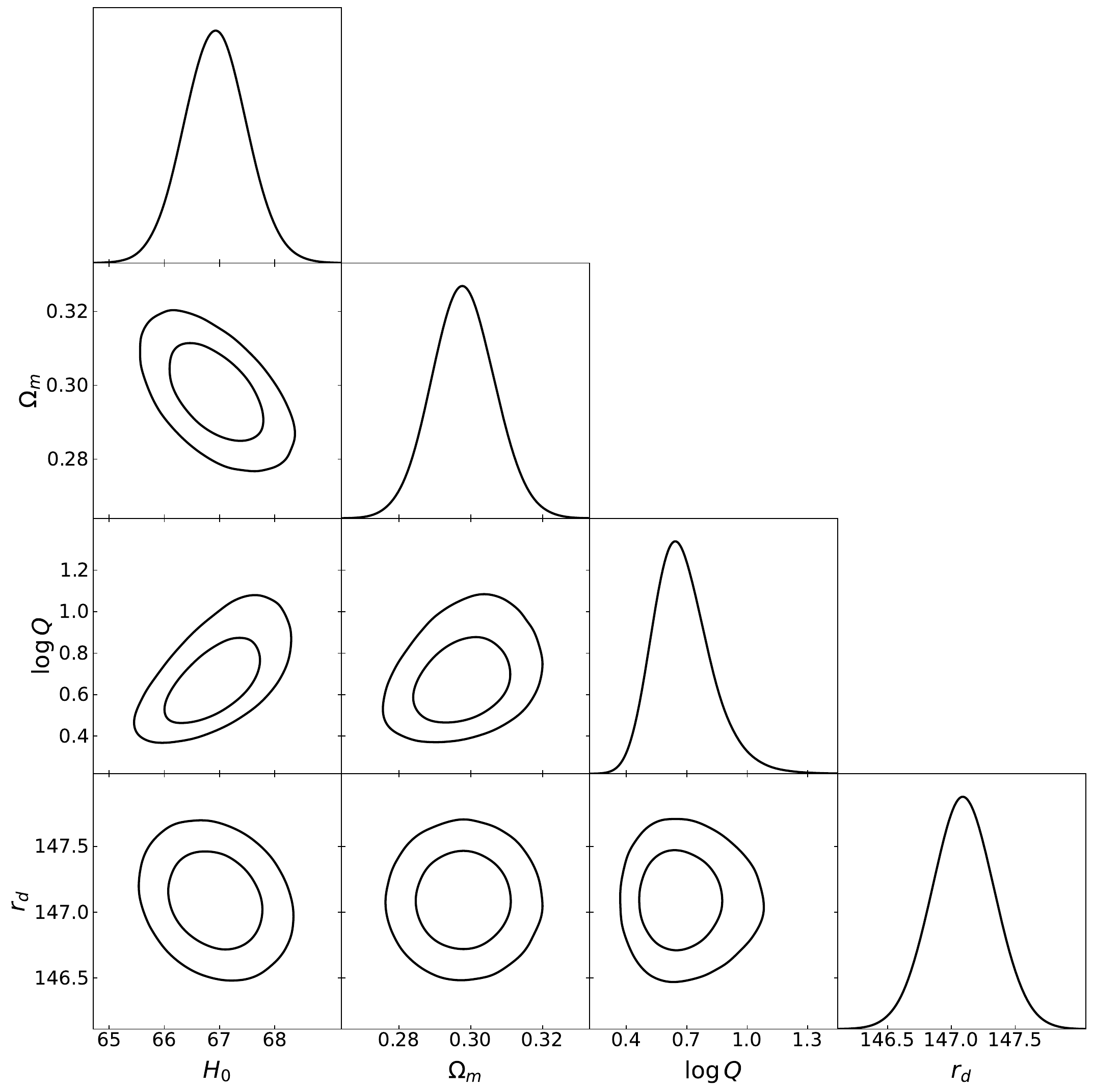}
  \end{center}
    \caption{
    One-dimensional posterior probability distributions and two-dimensional $68\%$ and $95\%$ confidence contours of the Hayward (left) and Bardeen (right) model parameters as inferred by the combination of dataset.
    }\label{fig1}
\end{figure*}

Since it is not possible to put upper bounds on the charge parameter of the Hayward, Bardeen and Asymptotic Safety CCRBH models, we only determine the lower bound and use log-uniform priors on the charge over the range $[e^{-20},e^5]$. On the other hand, we use flat prior on the charge of Dymnikova model in the range $[10^{-1}, 1.4]$. On the other parameters of the model, we use informative flat priors, i.e. for $H_0$ in the range $[40, 100]$ and $\Omega_{\rm m}$ in the range $[0.1, 0.5]$. The $1\sigma$ and $2\sigma$ contour plots are made with the package \texttt{GetDist} \cite{Lewis:2019xzd} and can be found in Figs.\ref{fig1} and \ref{fig2}.

\begin{table*}[t]
\centering

\begingroup

\setlength{\tabcolsep}{4pt} % Default value: 6pt
\renewcommand{\arraystretch}{1.7} % Default value: 1

\begin{tabular}{c|c|c|c|c|c}
\hline
Model & $\chi^2_{\rm tot,r}$ & AIC & BIC & $\Delta$AIC & $\Delta$BIC \\ \hline
$\Lambda$CDM & $1.032$ & $1786.80$ & $1803.16$ & $-$ & $-$ \\
$w_0w_a$CDM & $1.022$ & $1772.02$ & $1799.30$ & $-14.78$ & $-3.86$ \\
Hayward & $1.021$ & $1769.98$ & $1791.80$ & $-16.82$ & $-11.36$ \\
Bardeen & $1.022$ & $1770.16$ & $1791.99$ & $-16.64$ & $-11.17$ \\
Dymnikova & $1.021$ & $1769.89$ & $1791.71$ & $-16.91$ & $-11.45$ \\
AS & $1.022$ & $1770.67$ & $1792.49$ & $-16.13$ & $-10.67$ \\ \hline
\end{tabular}
\endgroup

\caption{Reduced $\chi^2$, model selection criteria and relative differences with respect to the $\Lambda$CDM baseline for the considered CCRBH DE models. }
\label{tab2}
\end{table*}

\subsection{Datasets}

We use some of the recent late-time cosmological data at the background level. In particular, the datasets employed are:
\begin{itemize}
    \item Baryon Acoustic Oscillation (BAO): The recent Dark Energy Spectroscopic Instrument (DESI) BAO measurements consists of the transverse comoving distance ($D_M/r_d$), the Hubble distance ($D_H/r_d$) and the angle averaged distance ($D_V/r_d$) \cite{Moon:2023jgl,DESI:2019jxc,DESI:2024mwx,DESI:2025zgx}. Here the distances are normalized to $r_d$, the comoving sound horizon at the drag epoch \cite{DESI:2025zgx}. These distances can be expressed in terms of the Hubble rate $H(z)$ as follows
    \begin{align}
        &D_M(z) = \frac{c}{H_0}\int_0^z\frac{dz'}{H(z')/H_0}, \\
        &D_H(z) = \frac{c}{H(z)}, \\
        &D_V(z) = (z D_M(z)^2D_H)^{1/3},
    \end{align}
    where $c$ is the speed of light in vacuum and we assumed a flat universe. We use the BAO results from DESI DR II based on the observations of the clustering of Bright Galaxy Samples (BGS), Luminous Red Galaxy Samples (LRG), Emission Line Galaxy (ELG), combined LRG and ELG, quasars and Lyman-$\alpha$ samples summarized in Tab.IV of Ref. \cite{DESI:2025zgx}. This dataset contains 13 data points spanning over the redshift range [0.1,4.16]. We use strong Gaussian prior on the comoving sound horizon at drag epoch $r_d$ from Planck data \cite{Planck:2018vyg} as $r_d = 147.09 \pm 0.26 {\rm Mpc}$. This dataset is referred to as `BAO'.

    \item Cosmic Chronometers (CC): The second dataset contains measurements of the expansion rate $H(z)$ of the universe from cosmic chronometers, in other words, they are differential ages of massive, early-time, passively evolving galaxies \cite{Jimenez:2001gg,Borghi:2021rft}. For our analysis, we adopt 15 data points reported in \cite{Moresco:2012by,Moresco:2015cya,Moresco:2016mzx} in the redshift range [0.1791, 1.965]. There are more than 30 CC data points available, but we choose only a subset of those where full subset of covariant matrix's non-diagonal terms and systematic contributions are reported in \cite{Moresco:2018xdr,Moresco:2020fbm}. We refer to this dataset as `CC'.

    \item Supernovae Type Ia (SNe Ia): The third dataset consists of observational data from type Ia supernovae. Specifically, we use the Pantheon+ compilation which contains 1701 spectroscopically confirmed SNe Ia in the redshift range [0.001, 2.26] \cite{Scolnic:2021amr}. The compilation provides the bias-corrected apparent magnitudes $m_{B,i}(z_i)$ together with the full statistical and systematic covariance matrix.  The theoretical distance modulus is defined as
    \begin{align}
        \mu_{th} = 5 \log_{10}\left(\frac{D_L}{{\rm Mpc}}\right) + 25, 
    \end{align}
    where $D_L = c(1+z)\int_0^z dz'/H(z')$ is the luminosity distance. The observed distance modulus is written as $\mu_i = m_{B,i} - {\rm M}$. Here ${\rm M}$ is a nuisance parameter which we analytically marginalize over using the procedure outlined in \cite{Bridle:2001zv,Scovacricchi:2015ely,Caroli:2021mjg}. We refer to this dataset as `SN'.
    
\end{itemize}

Combining these three cosmological observations, the final constraints on the model parameters are obtained by maximizing the total log-likelihood given by 
\begin{align}
    -2\log \mathcal{L}_{\rm tot} = \chi^2_{\rm tot}, 
\end{align}
where the total chi-squared $\chi^2_{\rm tot}$ is 
\begin{align}
    \chi^2_{\rm tot} = \chi^2_{\rm BAO} + \chi^2_{\rm CC} + \chi^2_{\rm SN},
\end{align}
and $\chi^2_{\rm BAO}$, $ \chi^2_{\rm CC}$ and $ \chi^2_{\rm SN}$ represent the chi-squared values of the respective datasets. We then compute 1D and 2D posterior probability distributions of the model parameters. We compare the performance of the considered models to flat $\Lambda$CDM using information criteria, in particular the Akaike Information Criterion (AIC) and the Bayesian Information Criterion (BIC) \cite{Akaike:1974vps,Schwarz:1978tpv,Trotta:2008qt}. They are given by
\begin{align}
    &{\rm AIC} = -2\log \mathcal{L}_{\rm max} + 2k, \\
    &{\rm BIC} = -2\log \mathcal{L}_{\rm max} + k \log N,
\end{align}
where $\mathcal{L}_{\rm max}$ is maximum likelihood, $k$ is the number of parameters and $N$ is the total number of data points. To assess the preference for a model, we compute the difference in information criteria (IC) between the considered model and $\Lambda$CDM, 
\begin{align}
    \Delta {\rm IC} = {\rm IC}_{\rm model} - {\rm IC}_{\Lambda CDM}. 
\end{align} 
In principle, a model with lower value of $\Delta {\rm IC}$ is preferred. A model with $\Delta {\rm IC} < 2$ is said to have strong support, $2<\Delta {\rm IC}<7$ suggests weak support and for models with $\Delta {\rm IC}> 10$ are strongly disfavored.

\subsection{Results}

\begin{figure*}[]
   \begin{center}
        \includegraphics[width=8cm]{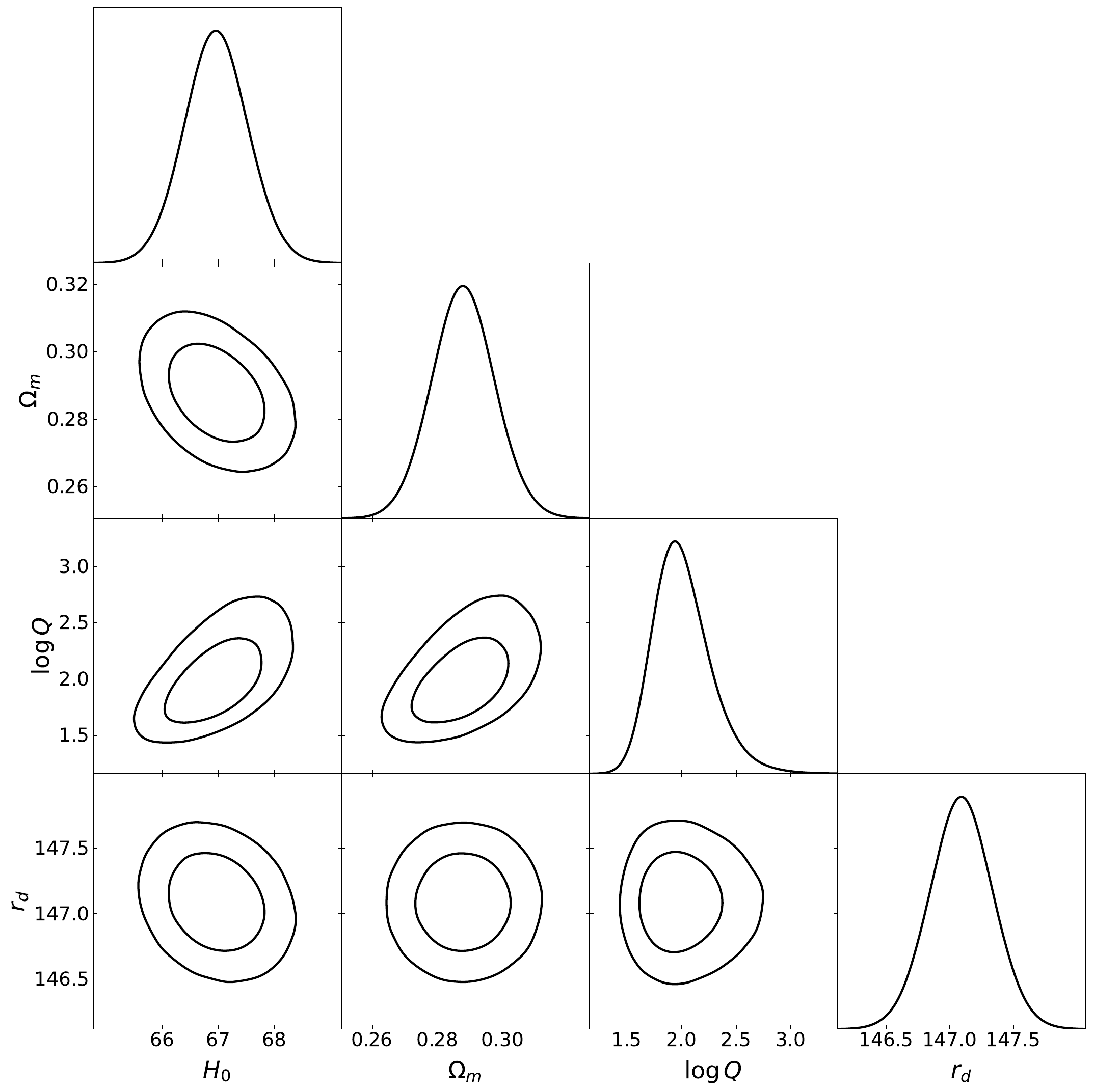}
        \includegraphics[width=8cm]{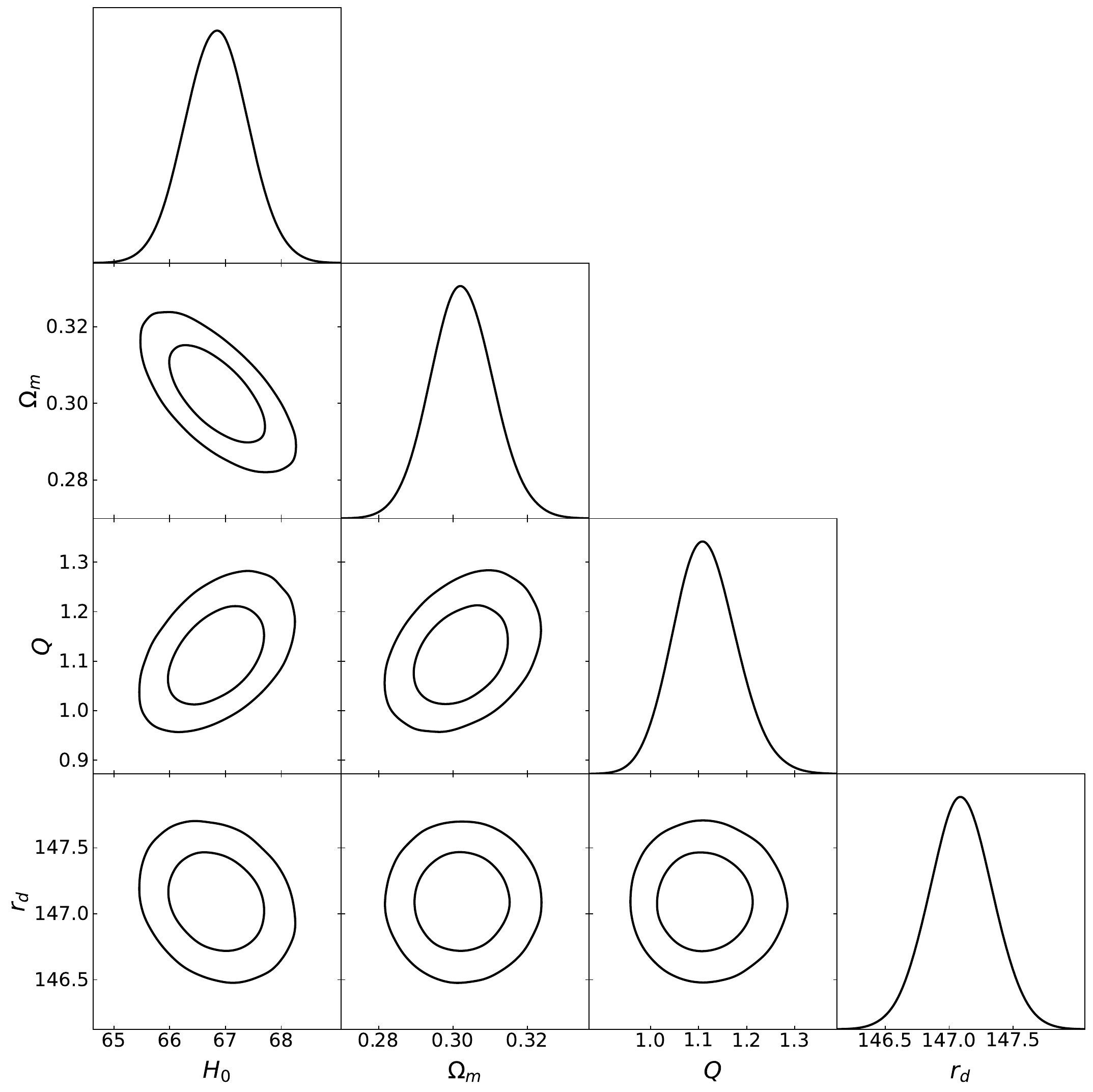}
  \end{center}
    \caption{
    One-dimensional posterior probability distributions and two-dimensional $68\%$ and $95\%$ confidence contours of the AS (left) and Dymnikova (right) model parameters as inferred by the combination of dataset.
    }\label{fig2}
\end{figure*}

We shall now discuss the results of the parameter inference procedure. In Table~\ref{tab1}, we present the median and 1$\sigma$ error bounds of the inferred parameter of $\Lambda$CDM, $w_0w_a$CDM and CCRBH DE models. Figures~\ref{fig1} and \ref{fig2} show one dimensional posterior probability distributions and two dimensional 68\% and 95\% confidence level contours for the free parameters of CCRBH DE models. 
From the above, we can see that all four CCRBH DE models agree well with the combination of the dataset considered for certain ranges of the charge values. As mentioned before, the charges of Hayward, Bardeen and Asymptotically Safe CCRBH DE models are unconstrained from above. This is a direct consequence of the fact that the dark energy component of these three models behaves similar to $\Lambda$ for higher values of the charge. Additional high redshift data is required to constrain the upper bound on the charge parameters and without it, the models become indistinguishable from $\Lambda$CDM. On the other hand, the charge parameter of Dymnikova CCRBH DE model is constrained from both ends. Note that the posteriors of the charges of Hayward, Bardeen and Asymptotically Safe CCRBH DE models look bounded on the corner plots because they are sampled in the log-space. For all the CCRBH DE models, we recover a lower value of $H_0$ because of the strong Gaussian priors imposed on $r_d$ and releasing the strong priors results in large error bars on $H_0$. 

In Table~\ref{tab2}, we report the information criteria for $\Lambda$CDM, $w_0w_a$CDM and CCRBH DE models for model comparison. All four CCRBH DE models perform better compared to $\Lambda$CDM for the considered combination of datasets in terms of both AIC and BIC. Although the CCRBH models give lower AIC and BIC values, this improvement originates mainly from the additional charge parameter. The data does not show a peaked posterior for this parameter for Hayward, Bardeen and AS, and yields only lower bounds. Hence the improvement in the information criteria does not correspond to a physical preference for the CCRBH DE models over $\Lambda$CDM. We also compare our modified dark energy models against the $w_0w_a$CDM model and although this parametrized model introduces two additional degrees of freedom, it does not yield a significant reduction in $\chi^2$ compared to our models. Our modified models achieve slightly lower AIC and BIC values than $w_0w_a$CDM, indicating that they provide a marginally better balance of goodness-of-fit and model complexity. However, the improvement remains modest, with $\Delta {\rm AIC} \le 3$ relative to $w_0w_a$CDM.

\section{Discussion} \label{sec5}

We investigated an hypothetical explanation for the late time accelerated expansion of the universe, as driven by UV modifications to GR in regular black hole solutions coupled to cosmic expansion. A generic feature of many regular back holes is the appearance of a de Sitter core near the center, regulated by a single UV cutoff parameter. The analysis of the available data on the universe's expansion rate allowed us to place constraints the black hole's parameter in order for the model to be a viable source of dark energy. A key question is then whether such models may be discarded from other, independent, observations. When it comes to black hole candidates in recent years there have been several kinds of observations, ranging from the shadow of the supermassive black hole candidates at the center of the Milky Way (known as Sagittarius-A* or Sgr-A*) and M87 galaxies \cite{Psaltis:2018xkc, EventHorizonTelescope:2019dse, EventHorizonTelescope:2020qrl, EventHorizonTelescope:2022wkp}, to the measurement of the electromagnetic spectrum of accretion disks surrounding supermassive black holes (see for example \cite{Bambi:2015kza, Tripathi:2018lhx, Cardenas-Avendano:2019zxd} and references therein), which can be used to place constraints on regular black hole charges \cite{Cadoni:2022vsn, Vagnozzi:2022moj}. Such constraints can in turn be used to test the DE hypothesis presented here, since its validity relies on the values of the charges to be consistent with those obtained from observations of black hole candidates.
It must be noted that the constraints on the values of the charges $q$ for the space-times considered here are obtained from the bounds on $Q$ that fit observations. Then for a sufficiently large boundary $r_b$ such constraints imply that the central objects must be horizonless. Such exotic objects, although highly hypothetical, have not yet been ruled out by astrophysical observations \cite{EventHorizonTelescope:2021dqv,EventHorizonTelescope:2022xqj,Vagnozzi:2022moj, Eichhorn:2022oma}.

Another important aspect to mention is that we expect almost all astrophysical black holes and compact objects to be rotating. However, analytical models for matching rotating black holes with cosmological solutions are not trivial and often require some kind of approximation \cite{Torres:2026jht}. At the same time, while the total mass of all the black holes in a given volume of the universe would be additive and sum up to the observed average density, one can expect the sum of the angular momenta to cancel out due to random spatial orientations. Therefore working with static regular black hole solutions appears to be a good assumption in this context.

At present, the available data is not enough to firmly rule out or validate the hypothesis that DE may be driven by the UV cutoff of regular black hole solutions. However, as future observations improve, we expect that tighter constraints on the geometry of black hole candidates such as Sgr-A* will allow to put tighter constraints on the validity of the hypothesis.

\section*{Acknowledgment}
The authors would like to thank Ernazar Abdikamalov for fruitful discussions. The authors acknowledge support from Nazarbayev University Faculty Development Competitive Research Grant Program No. 040225FD4737 `Modifications of General Relativity in the strong curvature regime and their implications for black holes and cosmology'. This research was funded by the Science Committee of the Ministry of Science and Higher Education of the Republic of Kazakhstan (Grant No. AP26103591).

\bibliography{ref.bib}

\end{document}